\documentclass[12pt]{article}
\usepackage{amsmath,latexsym}
\usepackage{graphicx}
\begin{document}

 \title{\Huge Some new  class of Chaplygin  Wormholes
  }
 \author{F.Rahaman$^*$, M.Kalam$^{\dag}$
 and K A Rahman$^*$}

\date{}
 \maketitle
 \begin{abstract}
Some new class of Chaplygin wormholes are investigated in the
framework of a Chaplygin gas with equation of state $ p = -
\frac{A}{\rho}$, $A>0$. Since empty spacetime ( $ p  = \rho = 0 $
) does not follow Chaplygin gas, so  the interior Chaplygin
wormhole solutions will never asymptotically flat. For this
reason, we have to match our interior wormhole solution with an
exterior vacuum solution i.e. Schwarzschild solution at some
junction interface, say $ r = a $. We also discuss the total
amount of matter characterized by Chaplygin gas that supplies fuel
to construct a wormhole.

\end{abstract}

  \footnotetext{
 $*$Dept.of Mathematics, Jadavpur University, Kolkata-700 032, India

                                  E-Mail:farook\_rahaman@yahoo.com\\
$\dag$Dept. of Phys. , Netaji Nagar College for Women, Regent Estate, Kolkata-700092, India.\\
E-Mail:mehedikalam@yahoo.co.in\\
}
    \mbox{} \hspace{.2in}

\title{ \underline{\textbf{Introduction}}: }
 Observations of Type IA supernovae and Cosmic microwave background
 anisotropy suggest that the Universe is currently undergoing an
 accelerated expansion [1-3]. After the publication of these
 observational reports, theoretical physicists have been starting
 to explain this astonishing phenomena theoretically. It is
 readily understand that this current cosmological state of the
 Universe requires exotic matter that produces large negative
 pressure. It is known as dark energy and it fails to obey Null
 Energy Condition (NEC). To describe theoretically this ghost like
 matter source, Cosmologists have been proposed several
 propositions as Cosmological constant [4]( the simplest and most
 popular candidate ), Quintessence [5]( a slowly evolving dynamical
 quantity which has a spatially inhomogeneous component of energy
 with negative pressure ), Dissipative matter fluid [6], Chaplygin
 gas[7]
 ( with generalized as well as modified forms ), Phantom energy [8-10](
 here, the equation of state of the form as $p = - w \rho $ with
 $ w  > 1$ ), Tracker field [11-12]( a new form of quintessence ) etc.
 Recently, scientific community shows great interest in wormhole physics because
 this opens a possibility to trip a very large distance in a very
 short time. In a pioneering work, Morris and Thorne [13] have shown
 that wormhole geometry could be found from Einstein general
 theory of relativity. But one has to tolerate the violation of
 NEC.

 That means the requirement of matter sources are the same as
 to explain the recent cosmological state ( accelerating phase )
 of the Universe. For this reason, Wormhole physicists have
 borrowed exotic matter sources from Cosmologists [14-23]. In this
 article, we will give our attention to Chaplygin gas as a
 supplier of energy to construct a wormhole. We choose Chaplygin
 gas as a matter source for the following reasons. In 1904, Prof.
 S. Chaplygin [24] had used matter source that obeys the equation
 of state as $ p = -\frac{A}{\rho}$, $A>0$ to describe elevating
 forces on a plain wing in aerodynamics process. This Chaplygin
 gas model has been supported different classes of observational
 tests such as supernovae data [25], gravitational lensing [26-27],
 gamma ray bursts [28], cosmic microwave background radiation
 [29]. The most important feature of Chaplygin gas is that the
 squared of sound velocity $v_s^2 = \frac{A}{\rho^2}$ is always
 positive irrespective of matter density ( i.e. this is always
 positive even in the case of exotic matter ). In the present
 investigation, we shall construct some new classes of wormholes
 in the framework of Chaplygin gas with equation of state $ p = -\frac{A}{\rho}$,
 $A>0$.
 For any matter distribution with spherical symmetry has no
 contribution at infinity i.e. then it is equivalent to empty
 space, in other words, the spacetime is a Schwarzschild
 spacetime. We note that the empty conditions i.e. $p=\rho=0$ are
 not feasible for Chaplygin gas equation of state. So, if one
 wishes to construct Chaplygin wormhole, then one has to match
 this interior wormhole solution with exterior Schwarzschild
 solution at some junction interface , say $r=a$. That means
 Chaplygin wormholes should not obey asymptotically flatness
 condition.

\title{ \underline{\textbf{Basic equations for constructing wormholes}}: }

A static spherically symmetric Lorentzian wormhole can be
described by a manifold $ R^2 X S^2 $ endowed with the general
metric in Schwarzschild co-ordinates $( t,r,\theta,\phi )$ as
\begin{equation}
                ds^2 = - e^{\nu} dt^2 + e^{\lambda} dr^2+r^2 d\Omega_2^2
            \label{Eq1}
          \end{equation}
Here, $\frac{\nu}{2}$ is redshift function   and

\begin{equation} e^{\lambda} = [1 - \frac{b(r)}{r}]^{-1} \end{equation}

where b(r) is shape function. The
  radial coordinate runs from $r_0$ to infinity, where the minimum  value $r_0$
  corresponds to the radius of the throat of the wormhole.
  Since, we are interested to investigate Chaplygin wormhole, so
  our wormhole spacetime will never asymptotically flat, in other
  words, we are compelled to consider a 'cut off' of  the stress
  energy  tensor at a junction interface, say, at $ r=a$.

\pagebreak

 Using the Einstein
field equations
 $G_{\mu\nu} = 8\pi T_{\mu\nu} $, in orthonormal reference frame
 ( with $ c = G = 1 $ ) , we obtain the following stress energy
scenario,

\begin{equation}\label{Eq5}
 e^{-\lambda}\left[ - \frac{1}{r^2}+\frac{\lambda^\prime}{r}\right]+\frac{1}{r^2}
 =  8\pi \rho
 \end{equation}

 \begin{equation}\label{Eq6}
  e^{-\lambda}\left[\frac{1}{r^2} +\frac{\gamma^\prime}{r}\right]-\frac{1}{r^2}
  =  8\pi p
 \end{equation}

\begin{equation}\label{Eq7}
  \frac{1}{2}e^{-\lambda}\left[\gamma^{\prime\prime} +
  \frac{1}{2}(\gamma^\prime)^2-\frac{1}{2}\gamma^\prime \lambda^\prime +
   \frac{\gamma^\prime -
  \lambda^\prime}{r}\right] = 8\pi p
 \end{equation}

where p(r) =   radial pressure = tangential pressure  ( i.e.
pressures are isotropic )
 and $\rho$ is the matter energy density.

 [`$\prime$' refers to differentiation  with respect to radial coordinate.]

The conservation of stress energy tensor $ [ T_a^b ]_ { ; b} =  0$
implies,

 \begin{equation}\label{Eq8}
  \frac{dp}{dr} =
  - ( p + \rho)\frac{\gamma\prime}{2}
  \end{equation}

Since our source is charaterized by Chaplygin gas, we assume the
equation of state as
\begin{equation}\label{Eq8}
 p = - \frac{A}{\rho}
  \end{equation}

Using equations (6) and  (7), one can get,

\begin{equation}\label{Eq8}
 \rho^2 = \frac{A}{1 + E e^{-\nu}}
  \end{equation}
where E is an integration constant.

Plugging equation (2) in (3) to yield

\begin{equation}
                \rho(r) =\frac{b^\prime}{8\pi r^2}
                \label{Eq2}
          \end{equation}
The solution of the equation $ b(r) = r $ gives the radius of the
throat $r_0$ of the wormhole. Now equation (6) gives the energy
density at $r_0$ as
\begin{equation}
\rho(r_0) = 8\pi r_0^2 A
\end{equation}

Since the shape function b(r) satisfies flaring out condition at
the throat ( i.e. $  b^{\prime} (r_0) < 1 $ ), then one can have
the following  restriction as

\begin{equation}
A < \frac{1}{(8\pi r_0^2)^2}
\end{equation}
Also violation of NEC, $ p + \rho < 0 $ implies
\begin{equation}
\rho < \sqrt{A}
\end{equation}
for all $ r \epsilon  [ r_0 , a ]$.

\title{ \underline{\textbf{Toy models of wormholes}}:

Now we consider several toy models of wormholes.

\textbf{Specialization one}:

Consider the specific form of redshift function as
\begin{equation}
               \nu =  2\ln E( 1+\frac{1}{r^3 })
            \label{Eq1}
          \end{equation}
where constant E is the same as in equation(8).

[ This function is asymptotically well behaved and always non zero
finite for all $r > 0$. So the choice is justified. ]
\begin{figure}[htbp]
    \centering
        \includegraphics[scale=.8]{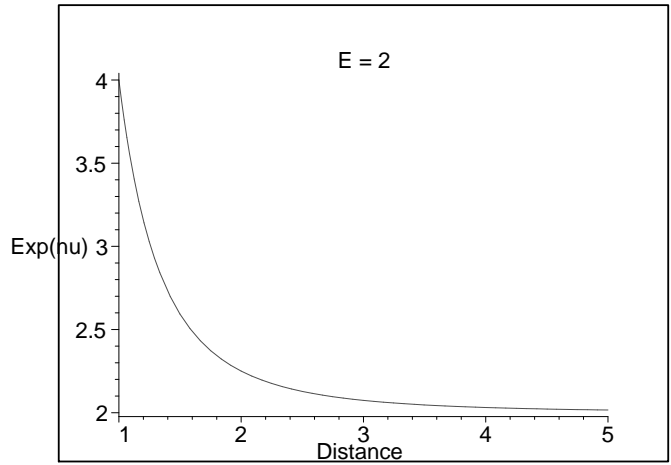}
    \caption{The variation of redshift function  with respect  to  r }
    \label{fig:wormhole}
\end{figure}

\pagebreak

 For the above redshift function (12),  we get an
expression for $\rho$ as
\begin{equation}
               \rho =  \left[ \frac{A(r^3 + 1)}{2r^3 +
               1}\right]^{\frac{1}{2}}
            \label{Eq1}
          \end{equation}
\begin{figure}[htbp]
    \centering
        \includegraphics[scale=.8]{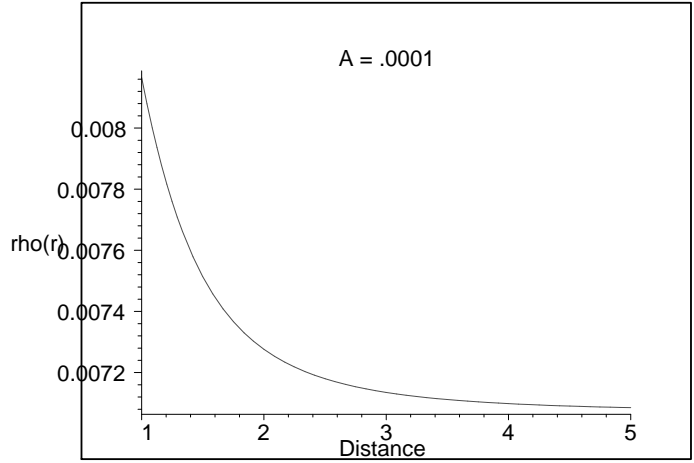}
    \caption{The variation of $\rho$  with respect  to  r }
    \label{fig:wormhole}
\end{figure}
 By using
(9), one can obtain the shape function as
\begin{equation}
                b(r)
                = \frac{4\pi\sqrt{A}}{3}[ \sqrt{2r^6 + 3r^3 + 1}]+ \frac{2\pi\sqrt{A}}{3}
                \ln[  \sqrt{2r^6 + 3r^3 + 1}+2r^3+ \frac{3}{2} ]
                \label{Eq10}
          \end{equation}
\begin{figure}[htbp]
    \centering
        \includegraphics[scale=.8]{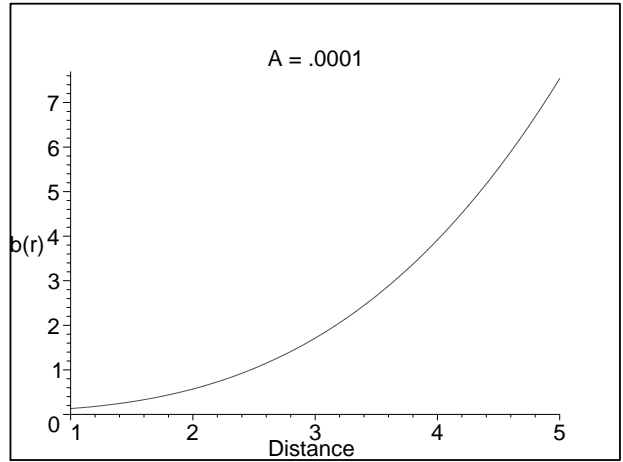}
    \caption{The variation of shape function  with respect  to  r }
    \label{fig:wormhole}
\end{figure}
We see that $ \frac{b(r)}{r} $ does not tend to zero as $ r
\rightarrow \infty $ i.e. the spacetime is not asymptotically flat
as expected for any Chaplygin wormhole spacetime. Here the throat
occurs at $ r = r_0 $ for which $ b(r_0)=  r_0 $  i.e. $r_0
                = \frac{4\pi\sqrt{A}}{3}[ \sqrt{2r_0^6 + 3r_0^3 + 1}]+
                \frac{2\pi\sqrt{A}}{3}
                \ln[  \sqrt{2r_0^6 + 3r_0^3 + 1}+2r_0^3+ \frac{3}{2}
                ]$.
                For the suitable choice of the parameter, the
                graph of the function  $ G(r) \equiv b(r)- r$ indicates
                the point $r_0$,  where G(r) cuts the 'r' axis (
                see fig. 4 ). From the graph, one can also note
                that when $ r> r_0 $, $ G(r) < 0 $ i.e. $ b(r)<
                r$. This implies $ \frac{b(r)}{r} <
                1$ when $ r> r_0 $.

\begin{figure}[htbp]
    \centering
        \includegraphics[scale=.8]{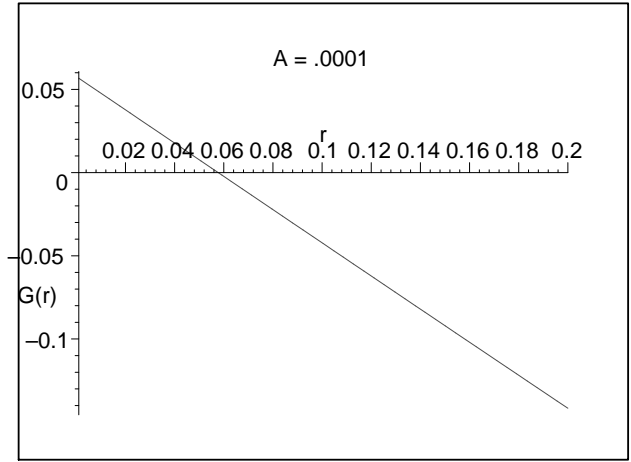}
    \caption{Throat occurs where G(r) cuts r axis }
    \label{fig:wormhole}
\end{figure}

\pagebreak

Now we match the interior wormhole metric to the exterior
Schwarzschild metric . To match the interior to the exterior, we
impose the continuity of the metric coefficients, $ g_{\mu\nu} $,
across a surface, S , i.e. $ {g_{\mu\nu}}_{(int)}|_S =
{g_{\mu\nu}}_{(ext)}|_S $.

[ This condition is not sufficient to different space times.
However, for space times with a good deal of symmetry ( here,
spherical symmetry ), one can use directly the field equations to
match [30-31] ]

 The wormhole metric is continuous from the throat, $ r = r_0$
to a finite distance $ r = a $. Now we impose the continuity of $
g_{tt} $ and $ g_{rr}$,

$ {g_{tt}}_{(int)}|_S =  {g_{tt}}_{(ext)}|_S $

$ {g_{rr}}_{(int)}|_S =  {g_{rr}}_{(ext)}|_S $

at $ r= a $ [ i.e.  on the surface S ] since $ g_{\theta\theta} $
and $ g_{\phi\phi}$ are already continuous. The continuity of the
metric then gives generally

$ {e^{\nu}}_{int}(a) = {e^{\nu}}_{ext}(a) $ and $
{g_{rr}}_{({int})}(a) = {g_{rr}}_{({ext})}(a) $.

Hence one can find

\begin{equation}e^{\nu}= ( 1 - \frac{2GM}{a}) \end{equation}
and $  1 - \frac{b(a)}{a} = ( 1 - \frac{2GM}{a})  $ i.e.
\begin{equation} b(a) = 2GM  \end{equation}

\pagebreak

Equation (16)  implies

$ a^3( E-1) + 2GMa^2 + E = 0 $

Thus matching occurs  at $ a =  S + T - \frac {a_1}{3}$,

 where $ S
= [ R + \sqrt{Q^3 + R^2}]^{\frac{1}{3}} $ and $  T = [ R -
\sqrt{Q^3 +
 R^2}]^{\frac{1}{3}}$,
$ Q = \frac { - a_1^2}{9},  R = \frac {-27a_3 -2a_1^3}{54}$

with $ a_1 = \frac{2GM}{E-1} $ , $ a_3 = \frac{E}{E-1} $.

One should note that the above equations (16) and (17) are
consistent equations if the solution of $'a'$ ( obtained from (16)
) should satisfy the equation (17). Thus one can see that the
arbitrary constants follow the constraint equation as

$2GM
                = \frac{4\pi\sqrt{A}}{3}[ \sqrt{2(S + T - \frac {a_1}{3})^6 +
                3(S + T - \frac {a_1}{3})^3 + 1}]
                \\
                +
                 \frac{2\pi\sqrt{A}}{3}
                \ln[  \sqrt{2(S + T - \frac {a_1}{3})^6 + 3(S + T - \frac {a_1}{3})^3 + 1}
                +2(S + T -
                \frac {a_1}{3})^3+ \frac{3}{2}
                ]$.

Now we investigate the total amount of averaged null energy
condition (ANEC) violating matter. According to Visser et al[32]
this can be quantified by the integral

\begin{equation}
                I = \oint ( p + \rho ) dV = 2\int_{r_0}^{\infty}( p + \rho
                )4\pi r^2 dr  \end{equation}

[ $ dV = r^2 \sin \theta dr d\theta d \phi $, factor two comes
from including both wormhole mouths ]

We consider the  wormhole field deviates from the throat out to a
radius $'a'$. Thus we obtain the total amount ANEC violating
matter as

$
 I_{total} =-\frac{4\pi\sqrt{A}}{3}\frac{ \sqrt{2a^6 + 3a^3 + 1}}{2}+
 \frac{2\pi\sqrt{A}}{\sqrt{2}}
                \ln[  2\sqrt{2}\sqrt{2a^6 + 3a^3 + 1}+4a^3+ 3 ]\\
                +\frac{4\pi\sqrt{A}}{3}\frac{ \sqrt{2r_0^6 + 3r_0^3
                + 1}}{2}- \frac{2\pi\sqrt{A}}{\sqrt{2}}
                \ln[  2\sqrt{2}\sqrt{2r_0^6 + 3r_0^3 + 1}+4r_0^3+ 3
                ]$

This implies that  the total amount of ANEC violating matter
depends on several parameters, namely, $a, A, G, M, E, r_0 $. If
we kept fixed the parameters $A, G, M, E, r_0$, then the parameter
$'a'$ plays significant role to reducing total amount of ANEC
violating matter. Thus total amount of ANEC violating matter can
be made small by taking suitable position, where interior wormhole
metric will match with exterior Schwarzschild metric. This would
be zero if one takes $ a \rightarrow r_0$.

\pagebreak

 \textbf{Specialization
two}:

Consider the specific form of shape function as
\begin{equation}
                b (r) =d \tanh Cr
            \label{Eq1}
          \end{equation}

where d and C ($ > 0 $) are arbitrary constants.

We will now verify the above particular choice of the shape
function would represent wormhole structure. One can easily that $
\frac{b(r)}{r} \rightarrow 0 $ as $ r \rightarrow \infty $ and
throat occurs at $r_0$ for which $ d \tanh Cr_0 = r_0 $. The graph
indicates the points $r_0$ where $ G(r) \equiv b(r) = r $ cuts
the r axis ( see fig. 6 ). Also from the graph, one can note that
when $r>r_0$ , $G(r) < 0$ i.e. $ \frac{b(r)}{r} < 1$ when $r>r_0$.

\begin{figure}[htbp]
    \centering
        \includegraphics[scale=.8]{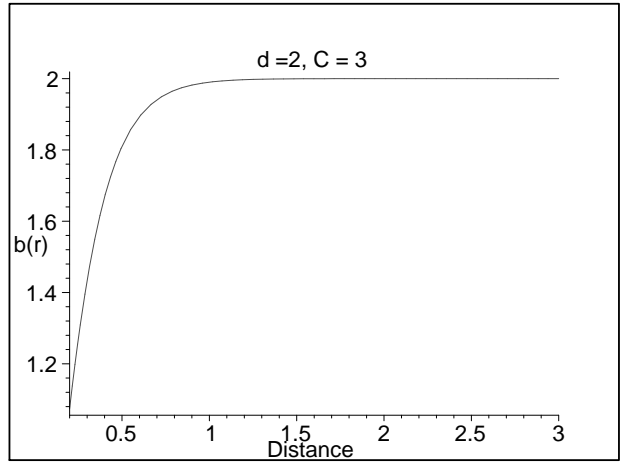}
        \caption{Diagram of the shape function of the wormhole}
   \label{fig:wh20}
\end{figure}

\begin{figure}[htbp]
    \centering
        \includegraphics[scale=.8]{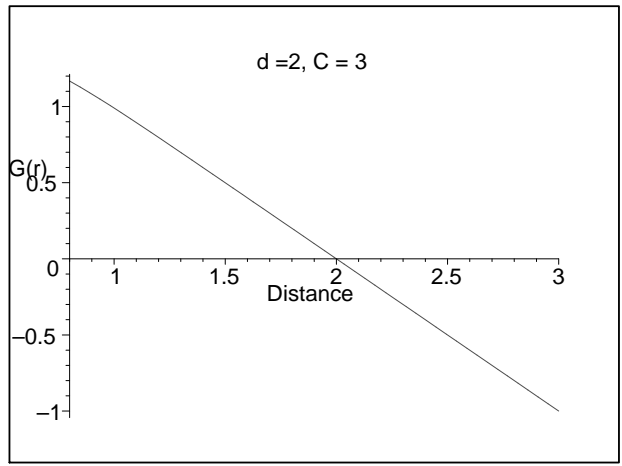}
    \caption{Throat occurs where G(r) cuts r axis }
    \label{fig:wormhole}
\end{figure}

\pagebreak

For this shape function, the energy density and redshift function
will take the following forms as

\begin{equation}
               \rho  = \frac {d C }{8 \pi r^2 \cosh^2 (Cr)  }
          \end{equation}

\begin{equation}
               e^\nu  = \frac {Ed^2 C^2 }{64A \pi^2 r^4 \cosh^4 (Cr) - d^2A^2  }
          \end{equation}

As the violation of NEC implies $ \rho^2 < A $, we see that
$e^\nu$ is regular in $[r_0, a]$, where $'a'$ is the cut off
radius.
\begin{figure}[htbp]
    \centering
        \includegraphics[scale=.8]{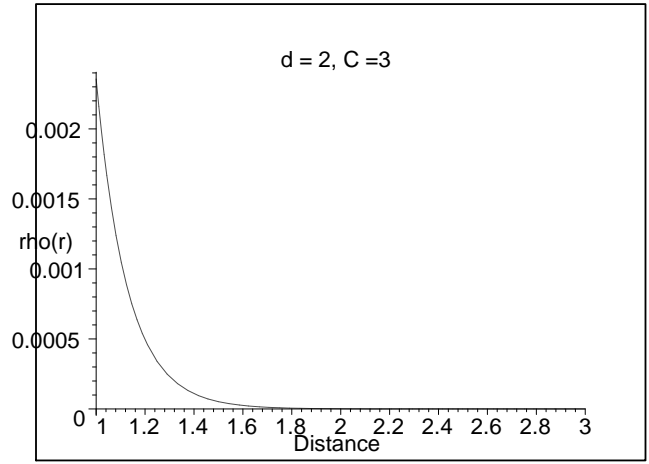}
    \caption{The variation of $\rho$  with respect  to  r }
    \label{fig:wormhole}
\end{figure}

\begin{figure}[htbp]
    \centering
        \includegraphics[scale=.8]{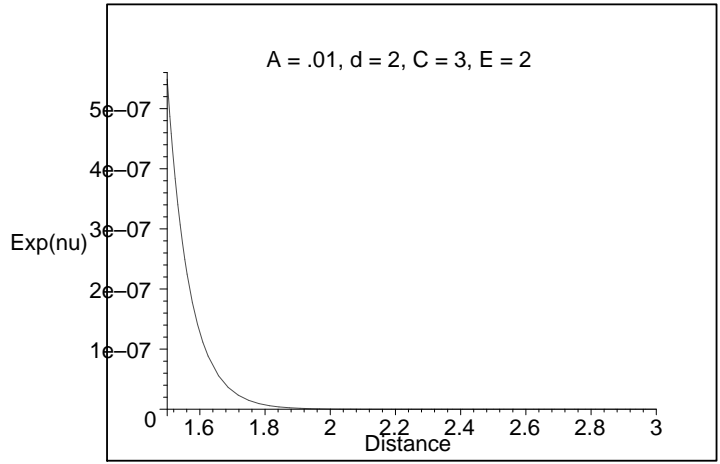}
    \caption{The variation of redshift function  with respect  to  r }
    \label{fig:wormhole}
\end{figure}

Since the wormhole metric is continuous from the throat $r = r_0$
to a finite distance $r=a$, one can use the continuity of the
metric coefficients across a surface, as above, i.e, at $'a'$.
Here we note that the matching occurs at $'a'$ where $'a'$
satisfies the following equation

\begin{equation}
             \alpha a^5 - \beta a^4 - \gamma - \delta = 0
          \end{equation}
where $\alpha = 64\pi^2 A$, $\beta = 128\pi^2GMA$, $\gamma =d^2(
1-4G^2M^2)^2(A^2+E^2C^2)$ and $\delta = 2GMd^2A^2( 1-4G^2M^2)^2$.

For this case, the total amount of ANEC violating matter in the
spacetime with a cut off of the strees energy at $'a'$ is given by

$
 I_{total} =\frac{d}{2}(\tanh Ca - \tanh Cr_0) -
 \frac{32\pi^2A}{dC}( \frac{a^5}{10} -
 \frac{r_0^5}{10})+\frac{1}{4C}( a^4 \sinh 2Ca - r_0^4 \sinh 2Cr_0)\\
- \frac{1}{2C^2}( a^3 \cosh 2Ca - r_0^3 \cosh 2Cr_0)
 - \frac{3}{4C^4}( a \cosh 2Ca - r_0 \cosh 2Cr_0)+\frac{3}{4C^3}( a^2
\sinh 2Ca - r_0^2 \sinh 2Cr_0)
 + \frac{3}{8C^5}(  \sinh 2Ca -
 \sinh 2Cr_0)$

In this case, also, if one treats C, $r_0$ , d, A are fixed
quantities, the total amount of ANEC violating matter can be
reduced by taking the suitable position where the interior
wormhole metric will match with exterior Schwarzschild metric.

 \textbf{Specialization
three}:

Now we make  specific choice for the  shape function as
\begin{equation}
                b (r) = D (1 -
\frac {F}{r} )(1 - \frac {B}{r} )
            \label{Eq1}
          \end{equation}

where F, B and D($ > 0 $) are arbitrary constants.

We note that $ \frac{b(r)}{r} \rightarrow 0 $ as $ r \rightarrow
\infty $ and throat occurs at $r_0$ for which $D (1 - \frac
{F}{r_0} )(1 - \frac {B}{r_0} ) = r_0 $. The graph indicates the
points $r_0$ where $ G(r) \equiv b(r) = r $ cuts the 'r' axis (
see fig. 10 ). Also from the graph, one can see that when $r>r_0$
, $G(r) < 0$ i.e. $ \frac{b(r)}{r} < 1$ when $r>r_0$.

\begin{figure}[htbp]
    \centering
        \includegraphics[scale=.8]{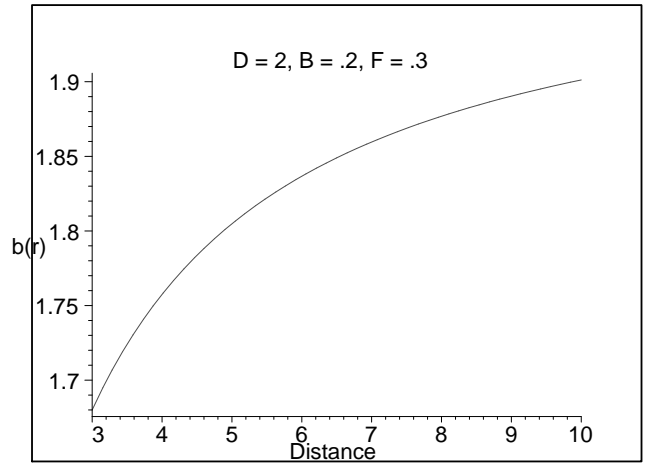}
        \caption{Diagram of the shape function of the wormhole}
   \label{fig:wh20}
\end{figure}

\begin{figure}[htbp]
    \centering
        \includegraphics[scale=.8]{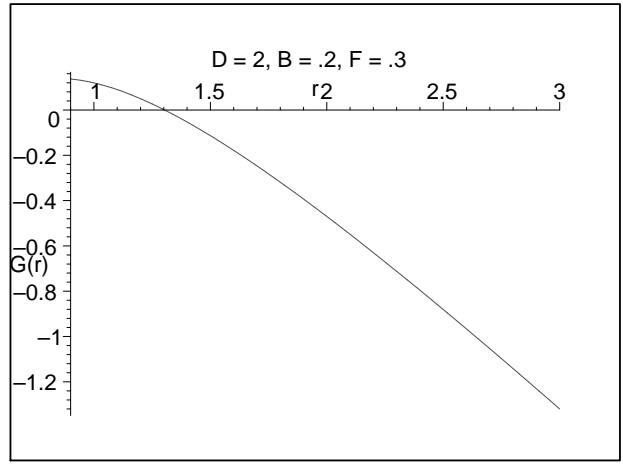}
    \caption{Throat occurs where G(r) cuts r axis }
    \label{fig:wormhole}
\end{figure}

For  the above shape function, we find the energy density and
redshift function as

\begin{equation}
               \rho  = \frac {D }{8 \pi r^4}[ ( F + B ) -
               \frac{2FB}{r}]
          \end{equation}

\begin{equation}
               e^\nu  = \frac {ED^2[ ( F + B ) -
               \frac{2FB}{r}]^2}{A (8\pi r^4)^2  - D^2[ ( F + B ) -
               \frac{2FB}{r}]^2 }
          \end{equation}

As above, the violation of NEC reflects that
 $e^\nu$ is regular in
$[r_0, a]$, where $'a'$ is the cut off radius.
\begin{figure}[htbp]
    \centering
        \includegraphics[scale=.8]{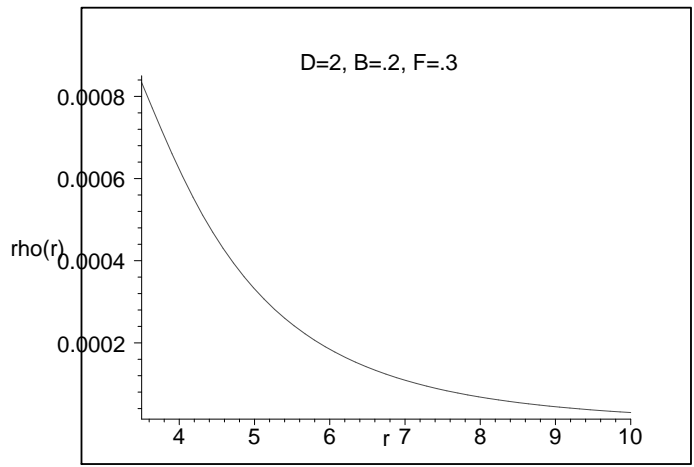}
    \caption{The variation of $\rho$  with respect  to  r }
    \label{fig:wormhole}
\end{figure}

\begin{figure}[htbp]
    \centering
        \includegraphics[scale=.8]{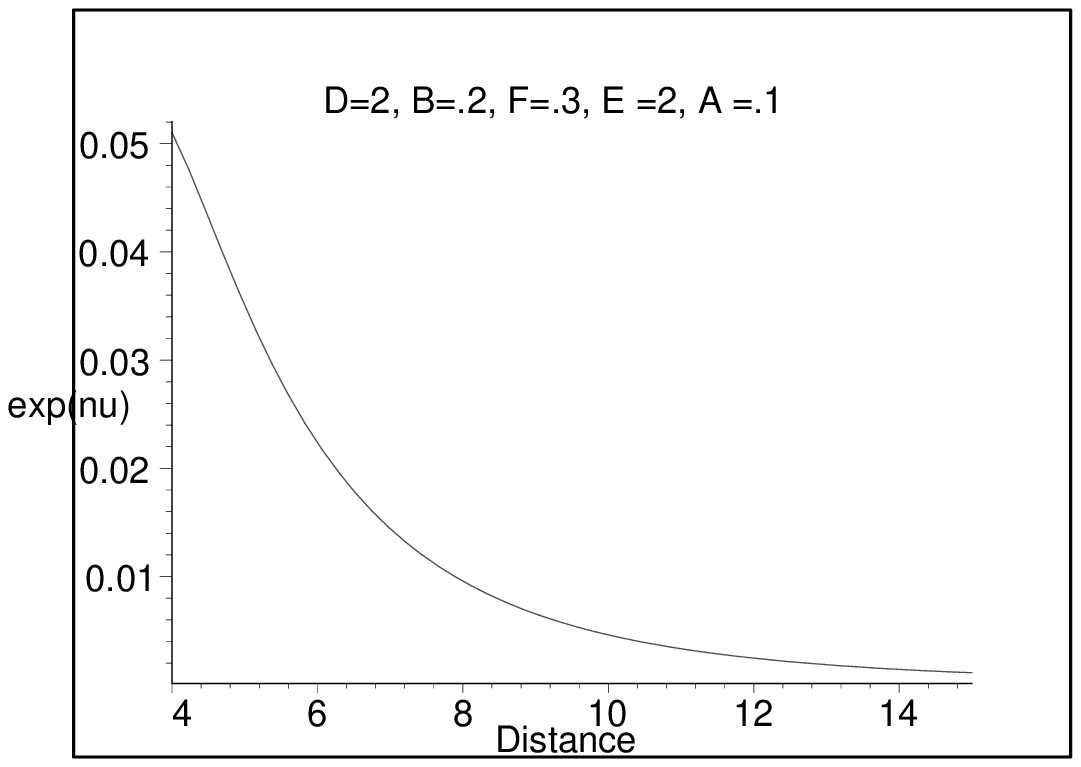}
    \caption{The variation of redshift function  with respect  to  r }
    \label{fig:wormhole}
\end{figure}

Now this interior wormhole  metric will match with exterior
Schwarzschild metric at $'a'$ where $'a'$ satisfies the following
system of consistent equations

\begin{equation}
                2GM = D (1 -
\frac {F}{a} )(1 - \frac {B}{a} )
            \label{Eq1}
          \end{equation}

\begin{equation}
              1 - \frac{2GM}{a}  = \frac {ED^2[ ( F + B ) -
               \frac{2FB}{a}]^2}{A (8\pi a^4)^2  - D^2[ ( F + B ) -
               \frac{2FB}{a}]^2 }
          \end{equation}

For this specific wormhole model, the total amount ANEC violating
matter in the spacetime with a cut off of the stress energy at
$'a'$ is

$
 I_{total} = 8DFB (  \frac{1}{a^2}- \frac{1}{r_0^2}) -
 \frac{D(F+B)}{2}(  \frac{1}{a}- \frac{1}{r_0}) -  \frac{32\pi^2A}{7D\alpha}(  a^7- r_0^7)
 - \frac{32\pi^2A\beta}{6D\alpha^2}(  a^6- r_0^6) +  \frac{32\pi^2A\beta^2}{5D\alpha^3}(  a^5- r_0^5)
  \frac{32\pi^2A\beta^3}{4D\alpha^4}(  a^4- r_0^4)+  \frac{32\pi^2A\beta^4}{3D\alpha^8}
  [ (\alpha a + \beta)^3- (\alpha r_0 + \beta)^3] -
  \frac{96\pi^2A\beta^5}{2D\alpha^8}[ (\alpha a + \beta)^2- (\alpha r_0 + \beta)^2]
  \frac{96\pi^2A\beta^6}{D\alpha^8}[ (\alpha a + \beta)- (\alpha r_0 + \beta)]
  - \frac{32\pi^2A\beta^7}{D\alpha^8}[ \ln(\alpha a + \beta)- \ln(\alpha r_0 +
  \beta)^2]$, where  $\alpha = F+B$ , $\beta =-2FB$.

Similar to previous cases,  if one treats C, $r_0$ , d, A are
fixed quantities, the total amount of ANEC violating matter can be
reduced by taking the suitable position where the interior
wormhole metric will match with exterior Schwarzschild metric.

\pagebreak

\title{ \underline{\textbf{Final Remarks}}: }

We have investigated how Chaplygin gas can fuel to construct a
wormhole. Since General Theory of Gravity admits wormhole solution
with the violation of NEC, so Scientists are trying to find matter
source that does not obey NEC. We give several specific toy models
of wormholes within the framework of Chaplygin gas. Since
Chaplygin wormholes are not asymptotically flat, so we have
matched our interior wormhole solution with exterior Schwarzschild
solution at some junction interface, say, $r=a$ [ Actually,  if
the metric  coefficients are not differentiable and affine
connections are not continuous at the junction then one has to use
the second fundamental forms associated with the two sides of the
junction surface ].  Though we have assumed several specific forms
of either redshift function or shape functions, but in all the
cases, one verifies the absence of event horizon. We also quantify
the total amount of ANEC violating matter for all models. One can
note that the total amount of ANEC violating matter depends on
several arbitrary parameters including the position of the
matching  surface. This position plays crucial role of reducing
matter. One can see that this would be infinitesimal small if one
takes $ a\rightarrow r_0$. Our models reveal the fact that one may
construct wormholes with arbitrary small amount of matter which is
characterized by Chaplygin equation of state. Finally we note that
the asymptotic wormhole mass ( defined by $ M = \lim _{r
\rightarrow \infty } \frac{1}{2} b(r)$ ) does not exist for first
model  but for the second and third models do exist and take the
values 'd' and 'D' respectively. In spite of these wormholes are
supported by the exotic matter characterized by  Chaplygin
equation of state, but asymptotic mass is positive. This implies
for an observer sitting at large distance could not distinguish
the gravitational nature between Wormhole and a compact mass 'M'.

 { \bf Acknowledgments }

          F.R is thankful to  DST , Government of India for providing
          financial support. MK has been partially supported by
          UGC,
          Government of India under MRP scheme. \\


\end{document}